\documentclass[a4paper,english,aps,manuscript]{revtex4}
\usepackage[T1]{fontenc}
\usepackage[latin9]{inputenc}
\usepackage{color}
\usepackage{rotfloat}
\usepackage{amstext}

\makeatletter

\@ifundefined{textcolor}{}
{%
 \definecolor{BLACK}{gray}{0}
 \definecolor{WHITE}{gray}{1}
 \definecolor{RED}{rgb}{1,0,0}
 \definecolor{GREEN}{rgb}{0,1,0}
 \definecolor{BLUE}{rgb}{0,0,1}
 \definecolor{CYAN}{cmyk}{1,0,0,0}
 \definecolor{MAGENTA}{cmyk}{0,1,0,0}
 \definecolor{YELLOW}{cmyk}{0,0,1,0}
}

\newcommand{\Htwop}{\hbox{\rm H}_2^+}
\newcommand{\Dtwop}{\hbox{\rm D}_2^+}

\newcommand{\HDp}{\hbox{\rm HD}^+}

\@ifundefined{showcaptionsetup}{}{%
 \PassOptionsToPackage{caption=false}{subfig}}
\usepackage{subfig}
\makeatother

\usepackage{babel}
\begin{document}

\title{Simple Molecules and Clocks}

\author{S. Schiller}

\affiliation{\textit{Institut für Experimentalphysik, Heinrich-Heine-Universität
Düsseldorf, 40225 Düsseldorf, Germany}}

\author{D. Bakalov }

\affiliation{\textit{Institute for Nuclear Research and Nuclear Energy, Tsarigradsko
chaussée 72, Sofia 1784, Bulgaria }}

\author{V.I. Korobov}

\affiliation{\textit{Joint Institute for Nuclear Research, 141980, Dubna, Russia}}
\begin{abstract}
The precise measurement of transition frequencies in cold, trapped
molecules has applications in fundamental physics, and extremely high
accuracies are desirable. We determine suitable candidates by considering
simple molecules with a single electron, for which the external-field
shift coefficients can be calculated with high precision.  Our calculations
show that $\Htwop$ exhibits particular transitions whose fractional
uncertainties may reach $2\times10^{-17}$ at room temperature. We
also generalize the method of composite frequencies, introducing tailored
linear combinations of individual transition frequencies that are
free of the major systematic shifts, independent of the strength of
the external perturbing fields. By applying this technique, the uncertainty
should be reduced to the $10^{-18}$ range for both $\Htwop$ and
$\HDp$. Thus, the theoretical results demonstrate that these molecules
are of metrological relevance for future studies.
\end{abstract}
\maketitle
\textit{Introduction}

Frequency metrology of cold trapped molecules is an emerging field
under intense development, driven by the promise of opening up new
but challenging opportunities in fundamental physics. It has been
proposed to use these systems to test the constancy of fundamental
constants related to particle masses, by measuring vibrational transition
frequencies over time or as a function of the gravitational potential
\cite{SS2-1}. Furthermore, the comparison of experimental molecular
transition frequencies with ab-initio results can be used to test
ab-initio theoretical calculations, in particular of QED effects \cite{SS3,Bressel 2012},
to measure mass ratios of small nuclei, and to search for a fifth
force on the sub-nanometer scale \cite{Salumbides}. The search for
parity violation effects on vibrational frequencies also requires
extreme frequency accuracy \cite{Chardonnet group}. A further potential
application is a test of Lorentz Invariance, using oriented molecules
\cite{Muller-1-1}. Different molecular systems, diatomic and polyatomic,
neutral and charged, are therefore being investigated \cite{Kotochigova 2009,Reinaudi 2012,Karr 2011}.

Concerning the constancy of the electron-to-nuclear mass ratio, microwave
cold atom clocks (exhibiting $2\times10^{-16}$ fractional uncertainty)
are already producing stringent limits. For molecules to become competitive
systems, they must therefore have a potential uncertainty in the $10^{-17}$
range. A crucial aspect in molecular frequency metrology is thus the
understanding of systematic frequency shifts in vibrational (or electronic)
transitions, the development of methods allowing their suppression
or, at least, their quantification, and the identification of candidate
systems \cite{Kajita YbLi analysis,Kajita and Minor CaH+ analysis,Zee2,kool,hfi,Bakalov and Schiller -quadrupole shift}.

In this paper, we discuss and answer affirmatively the question whether
it is in principle possible to reach extremely low inaccuracies ($10^{-18}$range)
in the measurement of transition frequencies of molecules. Our scenario
consists in considering \textit{simple} molecules, i.e. molecules
with one electron, for which the ab-initio theory has made significant
advances in the last decade \cite{Korobov 2012}. These allow not
only the ab-initio calculation of transition frequencies with fractional
inaccuracies of $4\times10^{-11}$, currently \cite{Korobov Hilico and Karr (2013)},
but also the accurate calculation of the sensitivity to external fields,
which is the focus here. Such ab-initio calculations were previously
performed for a few simple \textit{atomic} systems such as hydrogen
and one-electron highly charged ions \cite{Schiller HCI}.

A significant difference between atomic and molecular system is that
molecules have a multitude (many tens) of long-lived rovibrational
levels in their electronic ground state, each of which may have a
substantial number of hyperfine states. Thus, there is also a very
large number (e.g. thousands in molecular hydrogen ions) of transitions
having high spectroscopic quality factors. Their external-field shift
coefficients $\Delta\eta$ vary, often substantially, because the
states' rovibrational molecular wave functions vary and as a consequence
also the coefficients of the hyperfine Hamiltonian do. A subset of
these transitions may exhibit small external-field shifts. The computability
of the external-field shifts of simple molecules then has two main
consequences.  First, it permits selecting from this large set of
transitions the metrologically most advantageous ones (i.e. having
low sensitivity to external fields) based \textit{entirely on theory}.
Experimentally, one will apply the elegant techniques developed so
far in the field of atomic ion clocks, for measuring and minimizing
the various systematic shifts individually and estimating the residual
uncertainty.

Second, the computability also enables a new approach for a reduction
of the systematic shifts, which is particularly direct in molecules.
Here, one performs, in fairly rapid succession, measurements of a
set of $N$ selected transitions with frequencies $\{f_{1},\, f_{2},\ldots,\, f_{N}\}$
under time-constant and moderate, but otherwise arbitrary, external
perturbations $\{X_{j}\}$ , and numerically combines the results
with weights $\beta_{i}$ to a composite transition frequency $f_{c}=\sum_{i=1}^{N}\beta_{i}\, f_{i}$.
For the studies mentioned above, such a composite frequency is as
useful an observable as the frequency $f_{i}$ of an individual transition.

Consider now that each individual frequency $f_{i}$ is perturbed
by the external fields present (magnetic field, electric field, electric
field gradients, temperature, laser intensities, etc.) in a way expressible
as a power series, $f_{i}(\{X_{j}\})=f_{0,i}+\sum_{j}\Delta\eta_{j,i}(X_{j})^{n_{j}}$,
where $f_{0,i}$ are the unperturbed frequencies, and $\Delta\eta_{j,i}$
are the sensitivities to the external fields, given by the differences
of the sensitivities (shift coefficients) of the final and initial
states involved in the transition $f_{i}$, and calculable \textit{ab-initio}.
 Only those contributions that are relevant for a desired accuracy
of the composite frequency are included in the expansion, and the
possible occurrence of different powers $n_{j}$ for the same field
$X_{j}$ may also be taken into account.

The weights $\beta_{i}$ are computed from the conditions that the
sensitivities of the composite frequency to the external perturbations
(up to the orders described by the above power expansion) vanish:
$\partial f_{c}/\partial(X_{j})^{n_{j}}=\sum_{i=1}^{N}\beta_{i}\Delta\eta_{j.i}=0$.
If $M$ is the number of systematic effects to be canceled, including
different algebraic dependencies on the perturbation strengths, there
are $M$ such equations, and one needs to measure at least $N=M+1$
transitions, possibly having significantly different frequency, to
satisfy them. Thus, the $\beta_{i}$ are found by solving this set
of equations; we stress that the $\beta_{i}$ are functions of the
theoretical shift coefficients $\Delta\eta_{j,i}$, but are independent
of the external fields. At a simpler level, composite frequencies
are determined in atomic clocks, e.g. by averaging over several Zeeman
components of the same (clock) hyperfine transition in order to suppress
the linear Zeeman shift and electric quadrupole (EQ) shift \cite{EQ suppression in atomic ion clocks}.

Here we illustrate this concept for the one-electron molecules $\Htwop$
and $\HDp$; the extension to others, such as the other isotopologue
molecular ions $\Dtwop,\,\ldots$, is equally possible. Conceptually,
we envision the spectroscopy of these ions to be performed on a single
molecular ion, trapped in an ion trap. It is both sympathetically
cooled to the Lamb-Dicke confinement regime, and interrogated by a
laser-cooled atomic ion (Be\textsuperscript{+}) using a quantum-logic-type
\cite{Schmidt et al} or optical-force detection \cite{Optical force}.
Techniques of quantum-state preparation are applied \cite{Schneider rotational cooling,Saanum,Bressel 2012}.
We consider here only one-photon transitions, which avoid the relatively
large light shifts associated with the large intensities of the spectroscopy
laser in two-photon transitions \cite{Karr et al 2005,Bakalov and Schiller -quadrupole shift}.
In $\HDp$ the one-photon transitions are electric dipole (E1) transitions
with quality factors of order $10^{13}$; in $\Htwop$ one has to
resort to electric quadrupole (E2) transitions, since there are no
allowed E1 transitions in the ground electronic state. Such transitions
have been considered theoretically (without hyperfine structure effects)
in Refs.~\cite{Posen et al,Pilon and Baye-1}. Since the lifetime
of all $\Htwop$ levels exceeds $10^{6}\,$s, the transition quality
factor will in practice be determined by the laser line width or the
interrogation time. An electric quadrupole transition in a trapped
and cooled molecular ion has recently been observed \cite{Willitsch group}.

\textit{The systematic shifts and their calculation}

The external field shifts relevant for a trapped molecular ion are
the Zeeman shift, the Stark and EQ shift caused by the electric field
of the ion trap, the black-body radiation (BBR) shift, light shifts
and the 2\textsuperscript{nd}-order Doppler shift. In this work,
we treat explicitly the first four shifts. The light shift caused
by the spectroscopy laser can be made negligible, as is known from
research on atomic ion clocks that use E2 transitions. The 2\textsuperscript{nd}-order
Doppler shift scales inversely with the mass and thus will be significantly
larger than in typical atomic ion clocks, at the fractional level
$10^{-16}$, and its uncertainty is therefore a relevant issue. While
a discussion of the projected experimental level is beyond the scope
of this work, nevertheless a value in the $10^{-18}$ range might
be achievable.

We compute the systematic shifts by a combination of perturbation
theory and direct diagonalization, limiting the spin basis states
to those of a given level with vibrational and rotational quantum
numbers $v,\, L$. Where necessary we use highly accurate non-adiabatic,
variational wave functions \cite{Korobov - variational wavefunctions}.
Because the hyperfine splitting and Zeeman shift typically dominate
the other shifts, we first compute the eigenstates $|m(B)\rangle$
of the Hamiltonian $H_{{\rm eff}}^{{\rm hfs}}(v,\, L)+V^{{\rm mag}}(v,\, L)$
\cite{prl06,Zee2,KorPRA06R,Karr et al 2008}. The states $m$ are
labeled with $S$, the (approximate) quantum number of the total spin,
$J$, the total angular momentum, $J_{z}$, the projection on to the
$z$-axis parallel to the magnetic field ${\bf B}$, and, for $\Htwop$,
$I$, the quantum number of the total nuclear spin, or, for $\HDp$,
$F$, the quantum number of the electron-proton coupled spin. For
the Zeeman shift $E_{Z}(m(B))$, it is sufficient to consider the
first two terms, $E_{Z}(m)\simeq\eta_{B}B+\eta_{B^{2}}B^{2}$ \cite{Zee2,hfi}.
For each eigenstate, we then compute the expectation value of the
EQ and d.c.~Stark effective interaction Hamiltonian, $V^{\mathrm{EQ}}(v,\, L)+V^{\mathrm{S}}(v,\, L)$,
for given strengths of the additional external fields $X_{j}=V_{zz},\, E_{t},\, E_{z}$,
where $E_{t}$ $(E_{z})$ is the electric field component orthogonal
(parallel) to ${\bf B}$ and $V_{zz}=-\partial E_{z}/\partial z$.
 $V^{\mathrm{EQ}}$ and $V^{S}$ have been derived in \cite{Bakalov and Schiller -quadrupole shift,Bekbaev et al}
and only the results are given here. The EQ shift is, to a good approximation,
 $E_{EQ}(m)=$\-$\frac{3}{2}E_{14}(v,\, L)\, V_{zz}\langle m(B)|L_{z}^{2}-\mathbf{L}^{2}/3|m(B)\rangle$,
where the quadrupole coefficients $E_{14}(v,\, L)$ have been computed
in the Born-Oppenheimer (BO) approximation. The latter limits the
fractional accuracy to $\simeq10^{-3}$. Similarly, the Stark shift
of an energy level is $E_{S}(m)=$$-[\alpha^{(t)}(m(B))(E_{x}^{2}+E_{y}^{2})+\alpha^{(l)}(m(B))E_{z}^{2}]/2$,
where the transverse and longitudinal polarisabilities are computed
as $\alpha^{(t,l)}(m(B))=$$\alpha_{s}(v,\, L)+$$\beta^{(t,l)}\alpha_{t}(v,\, L)\langle m(B)|L_{z}^{2}-{\bf L}^{2}/3|m(B)\rangle$,
with $\beta^{(l)}=2$, $\beta^{(t)}=-1$. Note that the same matrix
element is involved in determining the dependency of the d.c.~Stark
and EQ shifts on the spin structure of a particular hyperfine state
$m$. We have obtained the scalar ($\alpha_{s}(v,\, L)$) and tensor
($\alpha_{t}(v,\, L)$) polarisabilities non-adiabatically, using
the non-relativistic variational wave functions, with inclusion of
only electric interactions. Details will be reported elsewhere \cite{Bekbaev et al}.
The inaccuracies of the polarisabilities stem from the neglect of
relativistic corrections (of relative order $\alpha^{2}$), and are
therefore of order $10^{-4}$ fractionally.

The BBR shifts of a transition, $\Delta f_{BB}$, are determined by
the dynamic polarisabilities, and for an isotropic (unpolarized) BBR
field only the scalar parts $\alpha_{s}(\omega)$ for the initial
and final states are relevant. The shift is to a very good approximation
independent of the hyperfine state and only depends on the rovibrational
levels $(v,\, L)$, $(v',\, L')$. With our extensive results on the
polarisabilities and accurate transition dipoles \cite{Tian et al}
we computed the BBR shifts and their temperature derivatives for relevant
transitions of $\HDp$, extending the results of Ref.~\cite{kool},
and of $\Htwop$. For the homonuclear ion $\Htwop$ the shift can
be approximately obtained from the static scalar polarisability only,
since its E1 transitions have much higher frequencies than the typical
BBR radiation frequencies, $\Delta f_{BB}(T_{0})=\Delta\eta_{T}\, T_{0}^{4}\simeq-(832\,{\rm V/m})^{2}(T_{0}/{\rm 300\, K)^{4}}\Delta\alpha_{s}/2h$,
where $T_{0}$ is the temperature of the BBR radiation field, and
$\Delta\alpha_{s}=\alpha_{s}(v',\, L')-\alpha_{s}(v,\, L)$. We computed
the correction of the shifts due to the frequency-dependence of the
contribution of the excited electronic levels to the polarisability
\cite{Bekbaev et al}. For $\Htwop$ we find it to be less than $1\times10^{-3}$
fractionally for the transitions $v=0,\, L=1\rightarrow v'<4,\, L'=1$
considered here.  Here, it is sufficient to use the static approximation
and thus the fractional inaccuracy of the BBR shift coefficients is
$\sigma_{y,\Delta\eta_{T}}=1\times10^{-3}$. For $\HDp$, when taking
into account all dynamic effects and neglect of relativistic corrections,
we reach a theoretical absolute uncertainty of the level BBR shifts
of 0.01~mHz. However, this uncertainty is so far available only for
a few levels and we therefore conservatively assume an uncertainty
$\sigma_{{\rm abs},\Delta f_{BB}}=0.1\,{\rm mHz}$.

\textit{Metrologically important transitions}

We have performed the analysis of the systematic shifts of $\Htwop$,
which we have evaluated for a large number of levels. Our computations
covered 26 states having $v$ up to $8$ and $L$ up to 4. We have
searched for metrologically advantageous transitions by first applying
the criterion of particularly small Zeeman shifts. For experimental
reasons it is reasonable to consider only transitions originating
in the vibrational ground state $v=0$ and we limited the final states
to those for which $v'\le4$. We also note that the E2 transition
strengths decrease rapidly with increasing $|v'-v|$ \cite{Posen et al}
and therefore small values are experimentally favorable. E2 transitions
with small linear Zeeman shifts are the pairs between homologous hyperfine
states, $I,\, S,\, J,\, J_{z}\rightarrow$ $I'=I,\, S'=S,\, J'=J,\, J'_{z}=J_{z}$
having $S=I+1/2$ when $I=1$. Their small linear Zeeman shift $\Delta\eta_{B}$
is a result of the near-cancellation of the shifts $\eta_{B}$ of
lower and upper state, which each lie in the range $|\eta_{B}|\simeq(0.15-1.5)$~MHz/G
\cite{Karr et al 2008}. Importantly, their average Zeeman shift vanishes.
A subset of favorable transitions is reported in Table \ref{tab:E2 rovibrational transitions}.
It presents, among the transitions with $|\Delta\eta_{B}|<10\,$Hz/G,
the 15 having the smallest absolute electric quadrupole shifts.

The spectroscopy of a single hyperfine transition can already reach
a high accuracy, for well-chosen transitions. We assume realistic
experimental conditions and performance \cite{Assumptions for single hyperfine transitions}
One technique for reducing some of the systematic shifts is based
on noting that, as in atoms \cite{EQ suppression in atomic ion clocks},
the electric quadrupole shift and the tensor polarisability of a state
are both proportional to $J(J+1)-3J_{z}^{2}$ in weak magnetic fields.
Therefore, both effects can be nulled by averaging over the $\Delta J_{z}=0$
Zeeman components of a transition $J,\, J_{z}\rightarrow J'=J,\, J_{z}'=J_{z}$,
where $J_{z}$ runs over all possible values $-J,\dots,J$. This approach
is only applicable to the $\Htwop$ case, where all such transitions
have small Zeeman shifts. The Zeeman shift is again nulled as well,
because of the equal and opposite shifts of the transitions chosen
above. The advantage compared to the orthogonal quantization technique
is an expected higher suppression factor of the EQ shift and the additional
nulling of the tensor Stark shift. Consider the 3.89~MHz hyperfine
line of $(v=0,\, L=2)\rightarrow(v'=1,\, L'=2)$. We introduce the
typical fractional time instability  $\sigma_{y,X_{j}}$ of the external
perturbation $X_{j}$ (i.e. $B,$ $E_{z},$ $E_{t}$, $V_{zz}$) on
the timescale of an individual transition frequency measurement, and
assume $(\sigma_{y,B},\,\sigma_{y,E},\,\sigma_{y,V_{zz}})=$ $(1,\,10,\,5)\times10^{-4}$.
We obtain the Zeeman shift and EQ shift uncertainties $(\sigma_{Z},\,\sigma_{EQ})/f=(4,\,4)\times10^{-18}$
and a negligible Stark shift uncertainty due to the field instability.
The scalar Stark shift is not nulled, and its absolute value, $1.1\times10^{-17}$
fractionally, is conservatively taken as Stark uncertainty $\sigma_{S}/f$.
The fractional BBR shift is unchanged by the averaging and is $\Delta f_{BB}/f=-9.7\times10^{-17}$.
While the theoretical uncertainty of this shift is negligible, the
fractional uncertainty $\sigma_{BB,T_{0}}$ associated with the experimental
uncertainty $\sigma_{T_{0}}$ of the BBR temperature is $\sigma_{BB,T_{0}}/f=$
$4(\sigma_{T_{0}}/T_{0})|\Delta f_{BB}|/f\simeq$ $1.0\times10^{-17}$.
The total uncertainty is $\sigma_{f,syst}/f=1.6\times10^{-17}$, an
outstandingly small value, similar to that of state-of-the art atomic
ion clocks.

In $\HDp$ we recently determined that transitions with zero total
angular momentum projection in the initial and final state, $J_{z}=0\rightarrow J'_{z}=0$,
are most favorable, since they exhibit a small quadratic Zeeman shift
at low field \cite{Bakalov and Schiller -quadrupole shift}. We found
no suitable transitions (within the reasonable requirement $v=0,\, v'\le5$)
having also particularly small electric quadrupole shift. We consider
the $J_{z}=0\rightarrow J_{z}'=0$ Zeeman component of the 71.1~MHz
hyperfine line of the $(v=0,\, L=3)\rightarrow(5,\,4)$ transition
(261~THz), with particularly small quadratic Zeeman shift ($\Delta\eta_{B^{2}}=-2.3$~Hz/G\textsuperscript{2}).
Conservatively, we take the residual Zeeman and Stark shifts as uncertainties.
Then the Zeeman, Stark and EQ shift uncertainties are $(\sigma_{Z},\sigma_{S},\sigma_{EQ})/f=$
$(0.4,\,1.7,\,28)\times10^{-17}$. The BBR shift is $\Delta f_{BB}/f=-1.8\times10^{-17}$,
and its theoretical uncertainty $\sigma_{BB,\eta_{T}}/f=2.6\times10^{-18}$.
The experimental uncertainty due to $\sigma_{T_{0}}$ is $\sigma_{BB,T_{0}}/f=5\times10^{-19}$
\cite{Bekbaev et al}. We see that the EQ shift uncertainty dominates
the total uncertainty, which at $\sigma_{f,syst}/f=3\times10^{-16}$
is significantly higher than for $\Htwop$.

\begin{table}
\subfloat{$\begin{array}{|c|c|cccc|cccc|r|c|r|c|c|c|}
\hline (\text{\textit{\ensuremath{v}}}\textit{'},\text{\textit{\ensuremath{L}}}\text{'}) & (\text{\textit{\ensuremath{v}}},\text{\textit{\ensuremath{L}}}) & I' & S' & J' & J_{z}' & I & S & J & J_{z} & \delta f_{0}\,\,\, & \Delta\eta_{B} & \Delta\eta_{V_{zz}}\,\,\,\, & \Delta\alpha^{(t)} & \Delta\alpha^{(l)} & \Delta f_{BB}\\
\text{upper} & \text{lower} &  &  &  &  &  &  &  &  & \text{[MHz]} & \text{[Hz/G]} & \text{\ensuremath{\mathrm{[{\rm Hz\,}m^{2}/{\rm GV]}}}} & \text{[at.\,\ u.]} & \text{[at.\,\ u.]} & {\rm [mHz}]\\
\hline \text{(1, 1)} & \text{(0, 1)} & 1 & \frac{3}{2} & \frac{5}{2} & \pm\frac{3}{2} & 1 & \frac{3}{2} & \frac{5}{2} & \pm\frac{3}{2} & -12.85 & \text{\ensuremath{\pm}4.20} & 3.2 & 0.75 & 0.69 & -6.3\\
\text{(1, 2)} & \text{(0, 2)} & 0 & \frac{1}{2} & \frac{5}{2} & \pm\frac{3}{2} & 0 & \frac{1}{2} & \frac{5}{2} & \pm\frac{3}{2} & -2.59 & \text{\ensuremath{\pm}8.61} & 4.5 & 0.77 & 0.67 & -6.4\\
\text{(2, 1)} & \text{(0, 1)} & 1 & \frac{3}{2} & \frac{5}{2} & \pm\frac{3}{2} & 1 & \frac{3}{2} & \frac{5}{2} & \pm\frac{3}{2} & -24.75 & \text{\ensuremath{\pm}9.24} & 6.6 & 1.71 & 1.55 & -14.3\\
\text{(1, 1)} & \text{(0, 1)} & 1 & \frac{3}{2} & \frac{5}{2} & \pm\frac{1}{2} & 1 & \frac{3}{2} & \frac{5}{2} & \pm\frac{1}{2} & -12.85 & \text{\ensuremath{\pm}1.40} & 12.6 & 0.82 & 0.55 & -6.3\\
\text{(1, 3)} & \text{(0, 3)} & 1 & \frac{3}{2} & \frac{3}{2} & \pm\frac{1}{2} & 1 & \frac{3}{2} & \frac{3}{2} & \pm\frac{1}{2} & -4.63 & \text{\ensuremath{\pm}6.02} & 12.7 & 0.84 & 0.56 & -6.4\\
\text{(1, 3)} & \text{(0, 3)} & 1 & \frac{3}{2} & \frac{9}{2} & \pm\frac{3}{2} & 1 & \frac{3}{2} & \frac{9}{2} & \pm\frac{3}{2} & -14.88 & \text{\ensuremath{\pm}7.56} & 13.4 & 0.84 & 0.55 & -6.4\\
\text{(1, 4)} & \text{(0, 4)} & 0 & \frac{1}{2} & \frac{9}{2} & \pm\frac{3}{2} & 0 & \frac{1}{2} & \frac{9}{2} & \pm\frac{3}{2} & -5.09 & \text{\ensuremath{\pm}9.87} & 14.6 & 0.87 & 0.54 & -6.5\\
\text{(1, 1)} & \text{(0, 1)} & 1 & \frac{3}{2} & \frac{5}{2} & \pm\frac{5}{2} & 1 & \frac{3}{2} & \frac{5}{2} & \pm\frac{5}{2} & -12.85 & \text{\ensuremath{\pm}7.00} & -15.7 & 0.62 & 0.96 & -6.3\\
\text{(1, 2)} & \text{(0, 2)} & 0 & \frac{1}{2} & \frac{3}{2} & \pm\frac{1}{2} & 0 & \frac{1}{2} & \frac{3}{2} & \pm\frac{1}{2} & 3.89 & \text{\ensuremath{\pm}4.27} & 15.9 & 0.85 & 0.51 & -6.4\\
\text{(1, 3)} & \text{(0, 3)} & 1 & \frac{3}{2} & \frac{9}{2} & \pm\frac{1}{2} & 1 & \frac{3}{2} & \frac{9}{2} & \pm\frac{1}{2} & -14.88 & \text{\ensuremath{\pm}2.52} & 17.7 & 0.88 & 0.49 & -6.4\\
\text{(1, 2)} & \text{(0, 2)} & 0 & \frac{1}{2} & \frac{5}{2} & \pm\frac{1}{2} & 0 & \frac{1}{2} & \frac{5}{2} & \pm\frac{1}{2} & -2.59 & \text{\ensuremath{\pm}2.87} & 18.2 & 0.87 & 0.48 & -6.4\\
\text{(1, 4)} & \text{(0, 4)} & 0 & \frac{1}{2} & \frac{7}{2} & \pm\frac{1}{2} & 0 & \frac{1}{2} & \frac{7}{2} & \pm\frac{1}{2} & 6.36 & \text{\ensuremath{\pm}4.13} & 19.1 & 0.9 & 0.48 & -6.5\\
\text{(1, 4)} & \text{(0, 4)} & 0 & \frac{1}{2} & \frac{9}{2} & \pm\frac{1}{2} & 0 & \frac{1}{2} & \frac{9}{2} & \pm\frac{1}{2} & -5.09 & \text{\ensuremath{\pm}3.29} & 19.5 & 0.9 & 0.47 & -6.5\\
\text{(2, 1)} & \text{(0, 1)} & 1 & \frac{3}{2} & \frac{5}{2} & \pm\frac{1}{2} & 1 & \frac{3}{2} & \frac{5}{2} & \pm\frac{1}{2} & -24.75 & \text{\ensuremath{\pm}3.08} & 26.2 & 1.87 & 1.23 & -14.3\\
\text{(2, 2)} & \text{(0, 2)} & 0 & \frac{1}{2} & \frac{3}{2} & \pm\frac{1}{2} & 0 & \frac{1}{2} & \frac{3}{2} & \pm\frac{1}{2} & 7.60 & \text{\ensuremath{\pm}9.24} & 33.0 & 1.94 & 1.14 & -14.4
\\\hline \end{array}$}\caption{\label{tab:E2 rovibrational transitions} Systematic shifts of selected
electric-quadrupole rovibrational transitions in $\Htwop$, ordered
according to the absolute value of electric quadrupole shift. The
transitions are between the levels $m:\,(v=0,\, L,\, I,\, S,\, J,\, J_{z})\rightarrow$
$m':\,(v',\, L',\, I',\, S',\, J',\, J_{z}')$ (lower $\rightarrow$
upper). Note that several Zeeman components of the same hyperfine
transition occur. $\delta f_{0}$ is the spin-dependent contribution
to the total transition frequency $f_{0}$, at 0~G. $\Delta\eta_{B}$
denotes the Zeeman shift coefficient of the transition frequency;
$\Delta\eta_{V_{zz}}$ is the electric quadrupole shift coefficient
at 0~G. $\Delta\alpha^{(t)}=\alpha^{(t)}(m')-\alpha^{(t)}(m)$, $\Delta\alpha^{(l)}=\alpha^{(l)}(m')-\alpha^{(l)}(m)$
are the transverse and longitudinal difference electric polarisabilities
betweeen upper ($m'$) and lower state ($m$), respectively, in atomic
units and in zero magnetic field. The two signs for $J_{z}$ and $J{}_{z}'$
indicate the transition pair $+\rightarrow+$,$-\rightarrow-$. $\Delta f_{BB}$
is the BBR shift at $T_{0}=300\,$K. For the $(0,\,1)\rightarrow(1,\,1)$
transition, the absolute frequency $f_{0}\simeq$65.6~THz, for the
$(0,\,1)\rightarrow(2,\,1)$ transition, $f_{0}\simeq$127~THz.}
\end{table}

\textit{The composite frequency method}

We now exemplify the concept of composite frequency which allows reducing
further the already small systematic shift uncertainties. A composite
frequency $f_{c}=\sum_{i}\beta_{i}\, f_{i}$ is free of Zeeman, quadrupole
and Stark shift if the four conditions $\sum_{i}\beta_{i}\Delta\eta_{B,i}=0$
(pure linear Zeeman effect for particular transitions of $\Htwop$),
$\sum_{i}\beta_{i}\Delta\eta_{B^{2},i}=0$ (pure quadratic Zeeman
effect for particular transitions of $\HDp$), $\sum_{i}\beta_{i}\Delta\eta_{V_{zz},i}=0$,
$\sum_{i}\beta_{i}\Delta\alpha_{i}^{(l)}=0$, $\sum_{i}\beta_{i}\Delta\alpha_{i}^{(t)}=0$
are satisfied, respectively, assuming that the potential individual
transitions are selected as described above. For the homonuclear molecular
hydrogen ions, the latter two conditions also eliminate the ``composite''
scalar polarisability and thus eliminate the (static) BBR shift, independently
of the temperature $T_{0}$, since all individual shifts are proportional
to $T_{0}^{4}$ in the static approximation. For $\HDp$ there is
no such simple dependence \cite{Bekbaev et al}, and the BBR shift
cancellation constraint, for a particular temperature $T_{0}$, is
$\sum_{i}\beta_{i}\Delta f_{BB,i}(T_{0})=0$, and represents a fifth
condition.

If we choose $N=5$ transitions for $\Htwop$ or $N=6$ for $\HDp$
we find a corresponding solution $\{\beta_{i}\}$ (up to a common
factor). But since there exists a large number $(\gg N)$ of transitions
with weak systematic shifts that may be employed, a large number $K$
of solutions $\{\beta_{i}\}_{K}$ exists, with a corresponding transition
set $\{i_{1},\ldots,\, i_{N}\}_{K}$ for each. We may therefore further
down-select the solutions according to additional criteria. Obviously,
the accuracy of cancellation of the shifts depends on the inaccuracies
of the theoretical shift coefficients $\sigma_{y,\eta_{j}}$ (which
as shown above are low and will be reduced further with future theory
work) and on the amount of variation $\sigma_{y,X_{j}}$ of the perturbations
in-between measurements of individual frequencies (which is to be
minimized experimentally).  We can then compute, for each solution,
the total absolute uncertainty $\sigma_{f_{c},syst}$ of the composite
frequency as $\sigma_{f_{c},syst}^{2}=$ $\sum_{j}\sigma_{j}^{2}=$
$\sum_{i,j}(\sigma_{y,\Delta\eta_{j,i}}^{2}+\sigma_{y,X_{j}}^{2}n_{j}{}^{2})\beta_{i}^{2}(\Delta\eta_{j,i}X_{j}^{n_{j}})^{2}$
and select a solution with a low value. Note that for given $\{\sigma_{y,\Delta\eta_{j,i}}\}$,
$\{\sigma_{y,X_{j}}\}$ and a desired level of $\sigma_{f_{c},syst}$
this leads to conditions for the maximum permitted strengths of the
external fields $X_{j}$.

We have performed a numerical search for the composite frequency with
lowest fractional systematic uncertainty $\sigma_{f_{c},syst}/f_{c}$.
We find that there are many solutions with very close values. Table
\ref{tab:Composite frequencies} gives one example for each ion.

% \begin{sidewaystable}
 \begin{table}
\caption{\footnotesize{ Examples of composite frequencies $f_{c}$
and the contributing individual transitions. Top: $\HDp$; bottom:
$\Htwop$. The uncertainties of the BBR shifts of the individual
transitions due to the BBR temperature uncertainty
$\sigma_{T_{0}}$, $\sigma_{BB,T_{0}}$= $\sigma_{T_{0}}|d\Delta
f_{BB}(T_{0})/dT_{0}|$ are listed; however, for both ions the
corresponding uncertainty of the composite BBR shift
$\text{\ensuremath{\Delta f}}_{BB,f_{c}}$ is negligible. $\Delta
f_{Z}$ is the Zeeman shift in 1~G. $\sigma_{Z}\text{,
}\sigma_{S,\text{transv}},\sigma_{S,\text{long}},\,\sigma_{\text{EQ}},\,\sigma_{BB}$
are the uncertainties of $f_{c}$ due to field instabilities and
theoretical shift coefficient uncertainties for, respectively, the
Zeeman, Stark (transv.), Stark (long.), EQ, and BBR effect. For
$\Htwop$, $B=0.1$~G,
$V_{zz}=0.67\times10^{8}\,$V/m\textsuperscript{2}, $\sigma_{y,B}=$
$1\times10^{-4}$, $\sigma_{y,\eta_{B},i}=0.1\times10^{-4}$,
$\sigma_{y,\eta_{T},i}=1\times10^{-3}$. For $\HDp$, $B=0.02$~G,
$V_{zz}=0.2\times10^{8}\,$V/m\textsuperscript{2}, $\sigma_{y,B}=$
$10\times10^{-4}$, $\sigma_{y,\eta_{B},i}=1\times10^{-4}$,
$\sigma_{{\rm abs},\eta_{T},i}=0.1\,{\rm mHz}$. Common parameters:
$T_{0}=300\,$K, $\sigma_{T_{0}}=8\,{\rm K}$,
$(\sigma_{y,E},\,\sigma_{y,V_{zz}},\,\sigma_{y,T_{BBR}})=$
$(10,\,1,\,5)\times10^{-4},$
$(\sigma_{y,\eta_{E},i},\,\sigma_{y,\eta_{V_{zz},i}})=$
$(3,\,1)\times10^{-4}$. The assumed uncertainty of the EQ shift
coefficient $\sigma_{y,\eta_{V_{zz}},i}$ is 10 times smaller than
available from our calcuations \cite{Bakalov and Schiller
-quadrupole shift}, but will be obtainable by using variational
rather than BO wavefunctions. Alternatively, $V_{zz}$ may be
reduced by a factor 10. Note the different values for$V_{zz}$
assumed in the calculation of the uncertainties for $\Htwop$ and
$\HDp$; however, for clarity of comparison, $\Delta f_{EQ}$ in
these tables is for a different, nominal value
$V_{zz,ref}=10^{8}$~V/m\textsuperscript{2}, for both ions.}}
\label{tab:Composite frequencies}
 \centering{}
 {\footnotesize{$\begin{array}{|c|c|cccc|cccc|r|rrrrrr|r|}
 \hline \text{HD}^{+} &
 \multicolumn{17}{c|}
 {f_{c}=54.04\, \text{THz,}\ \ \sigma_{f_{c},\text{syst}}/f_{c}=5.1 \times10^{-18},\ \
 \text{\ensuremath{\Delta f}}_{\text{BB},f_{c}}/f_{c}=\text{3.2} \ensuremath{\times10^{-17}}
 \text{}}
 %
% f_{c}= & 54.04 & \text{THz}, & \sigma_{f_{c},\text{syst}} &
% /f_{c} & =5.1 & \times10^{-18}, & \text{\ensuremath{\Delta f}}_{\text{BB},f_{c}} & /f_{c} & =
% & \text{3.2} & \ensuremath{\times10^{-17}} &  &  &  &  &
% \text{}
 %
 \\
 & \multicolumn{17}{c|}
 {(\sigma_{Z},\sigma_{S,\text{transv}},\sigma_{S,\text{long}},
  \sigma_{\text{EQ}}, \sigma_{\text{BB}})= (1.1,\,0.1,\,0.2,\,3.9,\,3.1)\times 10^{-18}}
%  &  &  &  &  &  & (\sigma_{Z}, & \sigma_{S,\text{transv}}, & \sigma_{S,\text{long}}, &
%  \sigma_{\text{EQ}}, & \sigma_{\text{BB}})= & (1.1,\,0.1, & 0.2,\,3.9,\, & 3.1)\times &
%  10^{-18} &  &  &
  \\
 \hline (v',L') & (v,L) & F' & S' & J' & J_{z}' & F & S & J & J_{z} &
 \text{\ensuremath{\delta f}}_{0} & \Delta f_{Z} & \Delta f_{\text{EQ}} &
 \Delta\alpha^{(t)} & \Delta\alpha^{(l)} & \text{\ensuremath{\Delta f}}_{\text{BB}} &
 \sigma_{BB,T_{0}} & \xi_{i}\\
 \text{upper} & \text{lower} & \multicolumn{4}{c|}{\text{upper}} & \multicolumn{4}{c|}{\text{lower}} &
 \text{[MHz]} & \text{[Hz]} & \text{[Hz]} & \text{[at.\,\ u.]} & \text{[at.\,\ u.]} &
 \text{[mHz]} & \text{[mHz]} & \text{}\\
 \hline
 \text{(1, 5)} & \text{(0, 4)} & 1 & 2 & 5 & 0 & 1 & 2 & 4 & 0 & -3.1 & -57.3 & -3.5 &
 \phantom{-0}0.4 & 1.6 & -9.1 & 0.7 & 1\phantom{.00}\\
 \text{(2, 4)} & \text{(0, 3)} & 0 & 1 & 5 & 0 & 0 & 1 & 4 & 0 & 30.7 & 31.7 & -4.0 &
 -\phantom{0}0.5 & 6.0 & -13.1 & 1.2 & -0.64\\
 \text{(2, 5)} & \text{(0, 4)} & 0 & 1 & 5 & 0 & 0 & 1 & 4 & 0 & 32.0 & -38.9 & -4.4 &
 \phantom{-0}1.1 & 2.8 & -13.5 & 1.3 & -0.75\\
 \text{(2, 5)} & \text{(0, 4)} & 0 & 1 & 6 & 0 & 0 & 1 & 5 & 0 & 31.2 & -39.6 & -4.0 &
 \phantom{-0}0.9 & 3.4 & -13.5 & 1.3 & -0.70\\
 \text{(3, 2)} & \text{(0, 1)} & 1 & 1 & 3 & 0 & 1 & 1 & 2 & 0 & -3.8 & 21.2 & -7.1 &
 -20.8 & 49.6 & -17.3 & 1.8 & 0.13\\
 \text{(5, 5)} & \text{(0, 4)} & 0 & 1 & 4 & 0 & 0 & 1 & 3 & 0 & 70.8 & 45.9 & -10.8 &
 \phantom{-0}8.3 & 1.4 & -37.3 & 4.1 & 0.40
 \\
 \hline\hline
 %\end{array}$}}{\small{\medskip{}}}
 %
 %{\footnotesize{$\begin{array}{|c|c|cccc|cccc|r|rrrrrr|r|}
 %\hline
 \text{H\ensuremath{_{2}^{+}}} &
% f_{c}= & 417.14 & \text{THz,} & \sigma_{f_{c},\text{syst}} & /f_{c} & =3.8 & \times10^{-18}, & \text{\ensuremath{\Delta f}}_{\text{BB,}f_{c}} & /f_{c} & = & 0 &  &  &  &  &  &
% \text{}\\
 \multicolumn{17}{c|}
 {f_{c}= 417.14 \, \text{THz,} \ \sigma_{f_{c},\text{syst}}/f_{c}=3.8\times10^{-18}, \
 \text{\ensuremath{\Delta f}}_{\text{BB,}f_{c}}/f_{c}= 0 \text{}}\\
 & \multicolumn{17}{c|}
 {(\sigma_{Z}\text{,} \sigma_{S,\text{transv}}, \sigma_{S,\text{long}}, \sigma_{\text{EQ}},
 \sigma_{\text{BB}})= (2.7,\,0.1,\,0.,\,2.4,\,0.9)\times10^{-18}}
 \\
 \hline (v',L') & (v,L) & I' & S' & J' & J_{z}' & I & S & J & J_{z} & \text{\ensuremath{\delta f}}_{0} & \Delta f_{Z} & \Delta f_{\text{EQ}} & \text{\ensuremath{\Delta\alpha}}^{(t)} & \Delta\alpha^{(l)} & \text{\ensuremath{\Delta f}}_{\text{BB}} & \sigma_{BB,T_{0}} &
 \xi_{i}\\
 \text{upper} & \text{lower} & \multicolumn{4}{c|}{\text{upper}}  & \multicolumn{4}{c|}{\text{lower}} & \text{[MHz]} & \text{[Hz]} & \text{[Hz]} & \text{[at.\,\ u.]} & \text{[at.\,\ u.]} & \text{[mHz]} & \text{[mHz]} &
 \text{}\\
 \hline \text{(3, 1)} & \text{(0, 1)} & 1 & \frac{3}{2} & \frac{5}{2} & \frac{5}{2} & 1 & \frac{3}{2} & \frac{5}{2} & \frac{5}{2} & -35.77 & 23.09 & -5.11 & 2.38 & 3.8 & -24.6 & 2.6 &
 1\phantom{.00}\\
 \text{(3, 1)} & \text{(0, 1)} & 1 & \frac{3}{2} & \frac{5}{2} & \frac{3}{2} & 1 & \frac{3}{2} & \frac{5}{2} & \frac{3}{2} & -35.77 & 13.86 & 1.02 & 2.95 & 2.7 & -24.6 & 2.6 &
 4.95\\
 \text{(1, 1)} & \text{(0, 1)} & 1 & \frac{3}{2} & \frac{5}{2} & \frac{3}{2} & 1 & \frac{3}{2} & \frac{5}{2} & \frac{3}{2} & -12.85 & 4.20 & 0.31 & 0.75 & 0.7 & -6.3 & 0.7 &
 -19.18\\
 \text{(1, 1)} & \text{(0, 1)} & 1 & \frac{3}{2} & \frac{5}{2} & \frac{5}{2} & 1 & \frac{3}{2} & \frac{5}{2} & \frac{5}{2} & -12.85 & 7.00 & -1.58 & 0.62 & 1.0 & -6.3 & 0.7 &
 -3.33\\
 \text{(1, 3)} & \text{(0, 3)} & 1 & \frac{3}{2} & \frac{3}{2} & -\frac{3}{2} & 1 & \frac{3}{2} & \frac{3}{2} & -\frac{3}{2} & -4.63 & -18.06 & -1.27 & 0.65 & 0.9 & -6.4 & 0.7 &
 -0.67
 \\
 \hline \end{array}$}}
 %\end{sidewaystable}
 \end{table}

For $\Htwop$ we have considered, in order to show the essence of
the method, the scenario where not all $J_{z}\rightarrow J_{z}'=J_{z}$
components of each hyperfine transition are measured, as in the example
above, but instead the minimum number $N=5$ of transitions that enables
canceling the $M=4$ systematic effects. In addition, the static BBR
shift is canceled ``for free''. The solution shown was chosen to
include only two values of $v'$, reducing the number of required
lasers to only two. The uncertainties of the scalar Stark shift and
of the BBR shift are now significantly reduced compared to a single
transition. In order to reduce the Zeeman shift uncertainty, first,
the Zeeman coefficient uncertainty $\sigma_{y,\eta_{B}}$ is assumed
very small, $10^{-5},$ which implies that relativistic contributions
need to be computed, which is feasible \cite{Korobov - relativistic polarizabilities}.
Second, a small magnetic field $B=0.1\,$G is chosen. This value is
still compatible with resolving individual Zeeman components, provided
appropriate ultra-narrow-linewidth lasers are employed. The BBR shift
uncertainty is due to the static approximation of the shift coefficients.
In the total uncertainty, $\sigma_{f_{c},syst}/f_{c}=4\times10^{-18},$
the Zeeman and EQ shift uncertainties are now the dominant ones.

In the case of $\HDp$, we show a particular solution where not the
BBR shift but its derivative with respect to temperature is canceled,
via the constraint $\sum_{i}\beta_{i}\, d\Delta f_{BB,i}(T_{0})/dT_{0}=0$.
We set again $B=0.02$~G but can relax the requirement for magnetic
field stability $\sigma_{y,B}$ compared to the $\Htwop$ case. This
results in a composite BBR shift $\Delta f_{BB,f_{c}}=$ $3\times10^{-17}$.
Its uncertainty $\sigma_{BB,\Delta\eta_{T}}/f_{c}=3\times10^{-18}$
is dominated by the theoretical uncertainties of the individual BBR
shifts. (As described above, this contribution will be reduced with
future theory work.) We find a total uncertainty $\sigma_{f_{c},syst}/f_{c}=5\times10^{-18}$,
limited in similar parts by the uncertainty of the EQ shift and of
the BBR shift. Alternatively, we can choose to cancel the BBR shift
instead of its derivative, but find that the best solutions yield
a 20\% larger total uncertainty.

\textit{\textcolor{black}{Extension of the method}}

With the proposed approach, additional systematic shifts can in principle
be compensated, as long as they are transition-dependent. Generally,
the introduction of additional conditions will require inclusion of
the same number of additional transitions in the composite frequency,
but in some cases, the multitude of solutions canceling $M$ shift
types allows selecting one that minimizes one additional shift type.
Compensation is possible, for example, of the light shift caused by
the UV laser that cools the atomic ion, and whose light field may
overlap with the molecular ion.  Along the same line, the small corrections
of the Zeeman shift contributions of higher order in $B$ may be eliminated,
which may be relevant when the applied magnetic field is not small
enough. For $\Htwop$ this is the quadratic contribution $\Delta\eta_{B^{2}}B^{2}$,
and for $\HDp$, $\Delta\eta_{B^{3}}B^{3}$, where the coefficients
are computable \cite{Zee2}. Therefore, the conditions $\sum_{i}\beta_{i}\Delta\eta_{B^{2},i}=0,$
or $\sum_{i}\beta_{i}\Delta\eta_{B^{3},i}=0$ can be added.

\textit{\textcolor{black}{In conclusion}}, we computed the external-field
shift coefficients of the one-electron molecular ions $\Htwop$ and
$\HDp$, and have identified vibrational transitions in $\Htwop$
having extremely low systematic shifts $(<2\times10^{-17})$. Moreover,
we have proposed to measure composite transition frequencies (selected
by theoretical calculation) that are free of external-field shifts
and should enable a systematic uncertainty as low as several $10^{-18}$.
The statistical uncertainty of the individually measured transition
frequencies would then be a significant contribution to the total
uncertainty, especially in the case of heteronuclear molecules, whose
natural transition Q-factors are limited to $10^{13}$. The expense
of the composite frequency method is the need of performing spectroscopy
and frequency measurements of $M+1$ transitions ($M$ being the number
of systematic effects to be canceled), in different wavelength ranges.
However, this is technologically feasible, as has been already shown
in the case of $\HDp$ \cite{SS3,Bressel 2012}. Thus, our theoretical
analysis provides a strong motivation and guide to future experiments
employing molecules to probe fundamental physics issues.
\begin{acknowledgments}
We thank H. Olivares Pilón for communicating results of unpublished
calculations.\end{acknowledgments}


\begin{thebibliography}{10}
\bibitem{Assumptions for single hyperfine transitions} Assumptions
are: (i) $\Htwop$:~$B=0.5\,$G (chosen so as to allow spectral separation
of the Zeeman components); $\HDp$: $B=0.02$~G; (ii) electric field
gradient $V_{zz}=0.67\times10^{8}\,{\rm V/m^{2}}$ . Note that $V_{zz}$
is determined by the trap endcap voltage and the axial potential cuvature
and both are large since the molecular ion and the atomic ion must
be close in order to apply the quantum logic technique. For $\HDp$:
averaging the transition over three orthogonal quantization directions
\cite{Bakalov and Schiller -quadrupole shift}, and achieving a reduction
of the EQ shift by a factor $100$; (iii) residual, uncalibrated trap
fields $E_{z}^{2}=E_{t}^{2}=(200\,{\rm V/m})^{2}$. The corresponding
Stark shift is taken as uncertainty; (iv) blackbody environment temperature
$T_{0}=300\,$K;  Its uncertainty is assumed as $\sigma_{T_{0}}=8\,$K.
\end{thebibliography}
\end{document}